\documentstyle[aasms4,12pt]{article}
\raggedright
\newcommand{\emo}{$\rm \mu _{814} (0)$}
\newcommand{\emu}{$\rm \mu _{814}$}

\newcommand{\mo}{$\rm \mu _{0}$}

\newcommand{\mss}{mag arcsec$^{-2}$}

\newcommand{\plm}{$\pm$ }
\newcommand{\lb}{$\langle \:$}
\newcommand{\gb}{$\rangle \:$}

\newcommand{\lta}{$\leq $}
\newcommand{\gta}{$\geq $}

\newcommand{\Bmo}{\rm ${\mu _B}$(0) }
\newcommand{\Imo}{\rm ${\mu _I}$(0) }

\newcommand{\alp}{$\alpha$\ }
\newcommand{\etal}{{\it et.al.}\ }
\newcommand{\degree}{$^{\circ}$ }

\newcommand{\Msol}{$M_{\odot}$\ }

\newcommand{\lsbgs}{\rm low surface brightness galaxies}

\input{epsf}
\input{rotate}
\begin{document}
\baselineskip = 10pt
\parskip = 10pt
\title{Structural Characteristics of Faint Galaxies Serendipitously Discovered
with the HST WFPC2}
\bigskip
\author{Karen O'Neil\altaffilmark{1}}
\bigskip
\affil{Arecibo Observatory, HC03 Box53995, Arecibo, PR 00612}
\medskip
\affil{email:koneil@naic.edu}
\medskip
\altaffiltext{1}{work done while at the University of Oregon}
\author{G.D. Bothun}
\medskip
\affil{Dept. of Physics, University of Oregon, Eugene OR, 97403}
\medskip
\affil{email:nuts@bigmoo.uoregon.edu}
\bigskip
\and
\bigskip
\author{C. D. Impey}
\medskip
\affil{Steward Observatory, University of Arizona, Tucson AZ, 85721}
\medskip
\affil{email:impey@as.arizona.edu}

\begin{abstract}
Utilizing the F814W and F300W filters, Hubble Space Telescope
Wide Field Planetary Camera-2 (WFPC2) images were taken of
four low surface brightness galaxies in
the direction of the Virgo cluster -- V7L3, V2L8, V1L4, and Malin 1.
The high resolution of the WFPC2 combined with the extremely diffuse nature of
the four galaxies makes them essentially transparent, allowing for
the serendipitous discovery of 139 background galaxies visible through both
the disks and nuclei of the foreground galaxies.
Surface photometry was done on the newly discovered galaxies through the
F814W (I-band) filter.  The detected galaxies have both r$^{1/4}$  and exponential
type profiles with radii (to the $\mu_{F814W}$ = 25.0 \mss\ limit) less than 5.0''.
Their total magnitudes range from 18.9 through the survey cut-off at 25.0
in the F814W filter.  The median central surface brightness
of those galaxies with exponential profiles is approximately 
one magnitude brighter than the background
F814W ``sky''.  Thus, with this dataset we recover Freeman's law and hence
know that we do not have a representative sample of distant galaxies
(and neither does anyone else).
When possible, the B, V, and I colors of these galaxies were determined
using ground-based images, which show the galaxies to be fairly red.
Coupled with their small angular size, we estimate the redshifts
to be 0.5 \lta\ z \lta\ 1.5.

Classification of the galaxies was done strictly in structural terms,
based only on the form of the derived luminosity profile.  No
morphological considerations were made during the classification process. 
23\% of the galaxies  we
detected have the $r^{1/4}$ profile typical of early type galaxies,
matching most previous studies of both the Hubble Deep Field
and the Medium Deep Survey which typically find 15\% $-$ 40\% E/S0 galaxies.
In addition, we have attempted to perform bulge/disk deconvolutions.
While we find that most of the sample cannot be easily deconvolved into
a classic bulge+disk, 7 objects could be fit in this way.  For these
7 objects we find a) a large range in bulge-to-total luminosity and
b) some disks which have a large bulge-to-disk ratio.   We also
present one object, 283-10, which is an excellent example of the
structural ambiguity that exists in the luminosity profiles of distant
galaxies.

In agreement with other studies we also found a significant
percentage of galaxies which have disturbed luminosity profiles indicative
of probable galaxy-galaxy interactions or mergers.  Indirect indicators
suggest that the volume over which $r^{1/4}$ objects are selected is
significantly larger than the volume over which disk galaxies are
selected.  This implies a relatively low space density of $r^{1/4}$
at all redshifts out to z $\sim$ 2.5 and is consistent with the general
idea that $r^{1/4}$ galaxies are largely confined to galaxy clusters.

\end{abstract}

\section{Introduction}

With the advent of deep ground based imaging, significant datasets
are being compiled that can address the issue of galaxy evolution.
While a representative sample of galaxies at any redshift has not
yet been obtained, there are by now several hundred galaxies with
redshifts available in distant galaxy surveys (e.g. Ellis \etal 1996;
Lilly \etal 1995; Cowie \etal 1996; Steidel \etal 1996; Cohen \etal 1996;
Morris \etal 1998).  These datasets include both galaxies in clusters
as well as the general ``field''.   Individual investigations of these
data sets are centered around three primary themes: 1) studying the
morphological mix of galaxies (e.g. ellipticals versus spirals) as
a function of redshift (see Marleau \& Simard 1998; Abraham \etal 1996);
2) studying the luminosity function of ``blue'' vs ``red'' galaxies
over various redshift ranges (e.g. Lilly \etal 1995; Hammer \etal 1997);
3) studying the evolution in the star formation rate as a function of
redshift (e.g. Madau \etal 1997; Meurer \etal 1997; Steidel \etal 1999).

In general, there is a fairly wide range of opinion on these matters
which reflects various levels of bias and incompleteness in the
available spectroscopic and photometric samples.  This
is to be expected as we are now just in the beginning phases of
data acquisition for distant galaxies and, after all, it took at least
50 years to recognize that our surveys of nearby galaxies are strongly
biased towards selecting mostly high surface brightness objects (see
Impey \& Bothun 1997 and references therein).  The level of surface
brightness and spectroscopic bias in  the selection of distant galaxies
is likely to be very complex and difficult to sort out (see Hogg  \etal
1999).    This, coupled with the difficulty of morphological classification of objects
only 1-2 arcseconds in diameter may be responsible for the current
divergence of opinion on the evolutionary rate of early type galaxies
and their possible dependence on environment. (see for instance,
 Barger \etal 1998; Balogh \etal 1999, von Dokkum \etal 1998;
Silva \& Bothun 1998; Schade \etal 1999).  As pointed out
succinctly by Andreon (1998), much of this disagreement may simply
be a reflection of morphological classification errors that cause
confusion or ambiguity in the E/SO classification.   
Such confusion can easily arise as 
the disk component of an SO galaxy is of lower surface brightness and
may be difficult to visually detect at small angular diameters.  
At the very least, it seems that the current results obtained by different
investigators are highly sample dependent and hence may not have general
applicability to galaxy evolution.

In this paper we try a different approach in characterizing the properties of
small, faint galaxies that are seen in the background of deep WFPC2
images obtained with HST.  Our approach, while considerably more
time consuming, relies entirely on the surface profiles of identified
galaxies for classification.  Unlike other investigations, which rely in
some measure on eyeball morphology (i.e. Driver, \etal 1995a, 1995b; Abraham, 
\etal 1996) no morphological considerations, based on visual
image inspection, are ever made during the {\it classification process}.
Instead,
once we decide an object has a high probability of being a distant
galaxy we
categorize it strictly on a structural basis depending on the form of
its luminosity profile.  In this way, a fairly explicit distinction
can be made between early type ($r^{1/4}$ profiles) and late type
($e^{-r/\alpha}$) galaxies in a way that is largely independent of
angular size (the galaxy just needs to be resolved by HST) and
surface brightness.   Indeed this approach, in principle, does allow 
for the detection of galaxies with large bulge-to-disk (B/D) ratios which
could, visually, be classified as pure ellipticals.  Examples of such
objects are found in this study.  Overall, our approach is
complementary to that of Simard \etal (1999) who study, to some
extent, the evolution in the bulge-to-total luminosity ratio.
Previous groups (e.g. Marleau \& Simard 1988, Ratnatunga \etal 1999)
have also used a similar model fitting approach.  In our treatment here,
however, we rely {\it only} on the
surface brightness decomposition for galaxy classification.

Our sample of background galaxies is a result of an interesting
consequence associated with observing foreground low surface brightness
(LSB) galaxies located in the Virgo cluster with the high angular
resolution of the HST WFPC2.
Basically, at 0.1 arcsecs per pixel, the HST WFPC2 doesn't really
detect a LSB galaxy but rather only records an elevated pattern
of noise (see O'Neil \etal 1999 for the explicit demonstration of
this).  As a result, the LSB is quite transparent (O'Neil \etal
1998) and background galaxies are readily detected, even when the LSB
takes up much of the WFPC2 frame.  
Although our sample is small, consisting of 139 identified galaxies in
the fields of 4 deep WFPC2 images, we are able to extract a variety of useful structural information.
Our fields are shown in Figure~\ref{fig:wfpc}
where it should be immediately obvious that its difficult to discern
the foreground LSB object, which in this case is typically an arcminute
in diameter.
These WFPC2 fields have been previously imaged using the
Las Campanas 2.5m Dupont
telescope (Impey, Bothun, \& Malin 1988; IBM hereafter) so when
identified on the WFPC2 images, we have some color information for
a subset of the galaxies.

In this paper we outline and apply our structural classification
scheme for analyzing distant galaxies.  Section 2 of this paper describes
the basic procedure of data reduction and identification of galaxies in
the WFPC2 data.  Section 3 presents the data and section 4 discusses
the properties of the galaxies we can infer.  Section 5 contains
a brief comparison of our results to other analysis of similar
kinds of data.  In an appendix we discuss the peculiarities
that are associated with some of the more unusual individual
galaxies we have detected.

\section{Data Reduction}

\subsection{Observations and Instrumentation}

All the data for this survey was taken using the HST WFPC2
on 1 May 1996, 3 August 1996, and 3 October 1996.  Each field was centered
around a known LSB galaxy in the direction of the
Virgo cluster and imaged through both the F814W (814) and
the F300W (300) filters.  The center of each LSB galaxy
is located in the WF3 field.
Figure~\ref{fig:wfpc} shows the full (mosaicked) images through the
814 filter, while the fields imaged are listed in Table~\ref{tab:hstf} --
the foreground galaxy name is given in column 1 while the coordinates of
the center of the field are given in column 2 and the observing dates are in column 3.
Each galaxy was imaged for a total of 2200 seconds (two 600s and two 500s images)
through each filter.

The WFPC2 consists of three Wide Field cameras and one Planetary camera.  The Wide Field
cameras have a focal ratio of f/12.9 and a field of
view of 80" x 80" with each pixel sub-tending 0.0996 arcsec$^2$.  The three cameras form an L-shape,
with the Planetary camera completing the square.  The Planetary camera
has a focal ratio of f/28.3,
0.0455 arcsec$^2$/pixel, and an overall field of view of 36 arcsec$^2$.  All four cameras have
an 800 x 800 pixel silicon CCD  with a thermo-electric cooler to suppress dark current.
The WFPC2 has two readouts formats -- single pixel resolution (FULL mode) and 2x2 pixel binning
(AREA mode).

Images were taken of each field through both the 814 and the 300
filter. The 814 filter is a broadband filter with $\lambda_0$ =
7924 \AA\ and $\Delta \lambda_{1/2}$ = 1497 \AA.   It is designed
to be similar to the Cousins I-band filter.  The 300 filter has
$\lambda_0$ = 2941 \AA\ and $\Delta \lambda_{1/2}$ = 757 \AA, and
is designed to be similar to the Johnson U-band filter.  Images
were for either 500s or 600s. The 814 images were all taken in
FULL mode, while the 300 images were all taken in AREA mode. As
the noise level through the 814 images was considerably lower then
through the 300 images, and most galaxies tended to be much
brighter through the 814, all galaxy identification was done with
the 814 images.  If a galaxy region could then be identified in
the 300 image it was analyzed in both colors. Otherwise all
information on the galaxy is from the 814 image.

The data reduction process, and calibration was performed  at STScI using the standard WFPC2-specific
calibration algorithms (the pipeline).  See the HST/WFPC2 Instrument Handbook for more
information about the calibration fields and procedures.

Four images were taken of each field, through each filter.  After the images were reduced,
they were inspected for obvious flaws such as filter ghosts or reflections.  If any flaws
existed in the frame an alternate frame was used and the offending frame was tossed.
Each frame was then shifted, registered and combined, using the STSDAS CRREJ ($\sigma$ = 10, 8, \& 6)
procedure to eliminate cosmic rays and
other small scale flaws.  The resultant images were then checked by eye to insure any registration
errors were under 0.5 pixel.

As few stars existed in the images,  a stellar point spread
function (PSF) was determined for each image using the Tiny Tim
software (Krist 1996; Remy, \etal 1997) with multiple wavelengths
(based on a F-type main sequence star), a base PSF of 3.0'' (1.5''
for the PC images), no sub-sampling, and no jitter correction. The
IRAF MEM routine was then used to deconvolve the stellar PSF from
the images, running it separately on each image and each chip.
As the PSF appeared to primarily affect only the inner 0.20'' of each 
galaxy (0.10'' on the PC chips), any galaxy which did not lie 
above the 25.0 814 \mss\ survey limit through a radii of at least 0.5'' 
(.25'' for the PC galaxies) was eliminated from our galaxy list, as 
such galaxies are not well resolved.

\subsection{Data Analysis}

The zeropoints for each field were taken from the PHOTFLAM value given in the image headers.
The zeropoint, in the STMAG system (the space telescope system based on a spectrum with
constant flux per unit wavelength set to approximate the Johnson system at V), is then
\[\rm ZP_{STMAG}\:=\:-2.5 log(PHOTFLAM)\:-\:21.1.\]
For the 814 filter, the PHOTFLAM was 2.5451 x 10$^{-18}$, corresponding to a zeropoint of
22.886, and for the 300 filter PHOTFLAM was
6.0240 x 10$^{-17}$, with a zeropoint of 19.450.  Due to filter variances, conversion
to the Johnson/Cousins U and I band was done using the conversion given in O'Neil,
\etal (1998) of 814 $-$ I = 1.43 \plm 0.05, 300 $-$ U = 0.04 \plm 0.05.
See Appendix A of O'Neil, \etal for more information on the magnitude systems
and conversion between the systems.

The peak intensity for each galaxy was found and ellipses were fit
around that point to obtain the intensity in each annulus using
the IRAF ELLIPSE routine. In the cases where no obvious peak
intensity existed in the galaxy (the more amorphous background
galaxies) the physical center of the galaxy, estimated by
centroiding with respect to outer isophotes, was chosen.  In the
cases of interacting (or overlayed) galaxies, the competing galaxy
was masked when possible, allowing for a surface brightness
profile to be obtained. On rare occasions, although two galaxies
appear to be merging, the merger appears close enough to being
complete that the galaxies were treated as one and a surface
brightness profile was fit to the entire object, with the center
having been chosen by the galaxies' peak intensity. 

The average,
sky-subtracted intensity within each (annular) ellipse was found
and calibrated with the photometric zeropoint.  The seeing (based
on the stellar PSF) gives a radius of 0.1" for the Planetary
camera (galaxies starting with 731, 281, m1, and 141), and 0.2"
for the Wide Field camera (all other galaxies), typical of WFPC2
data, inside of which the surface brightness profiles cannot be
trusted.  Surface brightness profiles were then plotted against
the major axis (in arcsec). Two different functions were then fit
against the deconvolved profiles -- an exponential profile and a
r$^{1/4}$-type profile.  The exponential profile is of the form
\begin{equation}\rm \Sigma (r)\:=\: \Sigma_0\:e^{-r \over \alpha} \label{eq:sige}\end{equation}
where $\Sigma_0$ is the central surface brightness of the disk in linear units
(\Msol/pc$^2$), and \alp is the exponential scale length in arcsec.  This can also be written
(the form used for data analysis) as
\begin{equation}\rm \mu (r)\:=\:\mu (0)\:+\:({1.086 \over \alpha})r  \label{eq:mue}\end{equation}
where $\mu(0)$ is the central surface brightness in \mss.  The r$^{1/4}$-type profile is
\begin{equation}\rm \Sigma (r)\:=\: \Sigma_0\:e^{({r \over
R_e})^{1/4}} \label{eq:r14} \label{eq:sigr}\end{equation}
where $\Sigma_0$ again is  the central surface brightness of the disk in linear units
and $R_e$ is the exponential scale length in arcsec.  In \mss, this equation reads
\begin{equation}\rm \mu (r)\:=\:\mu_e\:+\: 3.33 \left[\left({r \over R_e}\right)^{1/4} -1\right]
\label{eq:exp} \label{eq:mur}\end{equation}

where $\mu_e$ is the effective surface brightness in \mss\ and $\rm R_e$ is the
effective (1/2 light) radius in arcsec.   Profiles were fit to the data only between
r = 0.2'' (0.1'' for the PC data) and r$_{25}$, with the best fit profile being
determined by the data's $\chi^2$-fit to the profiles.

Isophotal, rather than aperture, annuli were fit to the galaxies for a 
number of reasons.  First, many of the galaxies in this study are not face on.
As such, aperture magnitudes do not accurately describe the light intensity
of the galaxy.  Second, most of the
galaxies do not lie in isolation in empty fields, but are in fact background 
objects behind foreground galaxies.  As such, it is not uncommon to have a
star, galaxy feature, or even another galaxy within close proximity of the 
studied object.  Since isophotal annuli follow the shape of the galaxy, accurate
information can be obtained to fairly high radii (major axis), while aperture
annuli will encompass the unwanted region more quickly.  If it were possible to
make a perfect mask of the unwanted region, this would not matter.  But since is
is extremely difficult to completely mask out all pixels affected by, say, a star,
aperture magnitudes often have to be restricted to small radii to avoid contamination
from nearby objects.  The results of the above two affects are this -- for an
inclined object, the isophotal and aperture magnitudes are different at small
radii and converge to the same value at high radii, while galaxies with nearby stars,
CCD flaws, etc. may never converge, as the large aperture annuli may me contaminated.

To offer a comparison between the two magnitudes (isophotal and aperture), 
Table~\ref{tab:apmag} provides the aperture magnitudes (in the F814W band)
and total isophotal magnitudes for all the galaxies in this study.  
Table~\ref{tab:apmag} is laid out as follows:\\
\hskip 0.2in {\bf Column 1:} The Galaxy name.\\
\hskip 0.2in {\bf Columns 2 -- 6:} Aperture magnitudes at radii of 0.5'', 1.0,'' 2.0 '',
3.0'' and 4.0''.\\
\hskip 0.2in {\bf Column 7:} The total isophotal magnitudes of each galaxy.\\
\hskip 0.2in {\bf Column 8:} The radius at the 25.0 \mss\ isophote, which is the same radii at\\
\hskip 0.25in which the isophotal magnitudes (Column 7) were determined.\\
Not surprisingly, the isophotal and annular values are within 0.1 magnitude for
the majority of the galaxies.  Of the 31 galaxies with magnitude differences
greater than 0.1 mag, 27 have inclinations greater then 60\degree.  The remaining
four either have inclinations of $\sim$55\degree and/or are extremely small
(r$_25$=0.5''), and none have isophotal/aperture differences greater than 0.2
magnitudes.  The isophotal magnitudes thus do give an accurate value for each
galaxies and are therefore used throughout the rest of this paper.

Galaxy structural types were assigned to three categories based strictly on
profile fit and not on any morphological criteria.  These categories are:\\
\hskip 0.5in $\bullet$ Class A:  Pure  r$^{1/4}$-type profile\\
\hskip 0.5in $\bullet$ Class B:  Good exponential fit with possible up or down turn in the
inner regions -- a subset of these galaxies will later be fit with
a combine Bulge + Disk model.\\
\hskip 0.5in $\bullet$ Class C:  No adequate fit to the profile\\
\vskip 0.1in

This classification system is similar to that used by Driver \etal
(1995a, 1995b), but we emphasize that these classifications were
based only on $\chi^2$-fit to the surface brightness profiles and
not on morphological inspection of the image.   For some $r^{1/4}$
profiles it is actually the fit to the outer part of the profile which
drives the  $\chi^2$ to its best value.  The very inner part of the
profile is sometimes not well fit.  This is mostly a problem for the
smallest galaxies where the deconvolution effects may still linger.

For the data in this survey the average sky brightness through the 814 filter was 23.0 \mss.
Galaxies with a central surface brightness as faint as 25.0 \mss\ (15\% of the sky background) were
detected, and an accurate (error \lta\  0.25 \mss) radial surface brightness profile
was typically found to 25.5 \mss\ (10\% of the sky background).

Galaxy inclination was found by using the IRAF ELLIPSE software to determine the major
and minor axis at each isophote.  The inclination angle is then
\begin{equation}\rm {\it i}\:=\:cos^{-1}{\left({r_{minor}\over r_{major}}\right)}.
\label{eq:inclin}\end{equation}
which we estimate to be accurate within \plm 5\degree.

\subsection{Galaxy Identification}

Once it was determined that a significant number of background
galaxies could be seen in the WFPC2 LSB galaxy images, an
intensive search was undertaken to identify as many background
galaxies as possible.  The search was undertaken by enlarging each
WF and PC image, within the IRAF environment, by a factor of four
and scanning the images by eye for non-stellar objects. By
examining both the 814 and 300 images available for each field a
minimum of four times, a list was compiled of all possible
non-stellar objects which had a minimum radius of 5 pixels.
All objects on the list then had their appearance checked against
their image in one of the uncombined frames to insure no errors
had occurred during the image processing phase (e.g. image
registration errors).  Remaining objects were considered potential
galaxies and left on the list.  It should be noted that because of
the high noise in the 300 frames, all galaxy identification was
ultimately done in the 814 frames.  Many of the galaxies could
not even be found in the 300 frames even after being
clearly identified in the 814 frames.

We did attempt to utilize the FOCAS software to examine each image for objects
at least 4 pixels in radius and at least 2$\sigma$ above
the sky background.   Unfortunately, though, FOCAS proved remarkably inept at identifying the
majority of the background galaxies.
Many of the galaxies found in the survey fields have a
non-spherical, amorphous appearance.   If the galaxy was bright enough FOCAS {\it usually}
identified the object as a potential galaxy candidate.  As the galaxies became fainter, though,
FOCAS relied on the objects having a core with contiguous pixels brighter than the
given threshold (2$\sigma$, typically)  and usually missed both the fainter and the more interesting background
galaxies (e.g. 284-15).  Thus although our `by eye' method of searching for galaxies is more
tedious than the usual automatic scanning methods, the non-conventional appearance of
these galaxies made our method more accurate in detecting all the objects to a
central surface brightness of approximately 25.0 \mss\ (10-15\% of the sky brightness),
giving us a more complete list of galaxies in the studied regions than would otherwise have
been possible.  The difficulty FOCAS had in identifying galaxies which were
either very faint or `non-conventional' in appearance has also been shown
to hold true in nearby (z\lta\ 0.02) LSB galaxy searches, for
identical reasons (see O'Neil 1997; O'Neil, Bothun, \& Cornell 1997a; O'Neil, \etal 1997b).
This has significant implications for the yield of LSB objects found by
the Sloan Digital Sky Survey (SDSS) if reliance is made solely on automatic
image detection software.

After the initial list of potential galaxies was compiled, all the candidates were visually inspected
to determine the likelihood that they are true background galaxies and not a part
of the foreground LSB galaxy or random noise.  For example,
discrete blobs within an LSB galaxy do not generally have
exponential or r$^{1/4}$ profiles.  This led to a set of visually
qualitative indicators
used to assess the probability that a faint image was indeed that of
a distant galaxy.
 Each galaxy's structural appearance,
surface brightness profile, and location within the frame was examined, and a rating
was given to the galaxy in each category.

For structural appearance, the galaxy was given
a 3 if it looked like a typical galaxy, a 1 if it had a completely unconventional appearance,
and a rating of 2 if it lay in between.  The surface brightness profiles of the
galaxies were examined similarly.  If the galaxy had a nice exponential or r$^{1/4}$-type
of profile, it was given a rating of 3.  If its profile was fairly close to exponential
or  r$^{1/4}$, or if it had a rounded profile, it was given a rating of
2.  If the profile was noisy, or contained a number of `bumps', it was given a rating
of 1.  Finally, if the surface brightness profile of the galaxy appeared to simply
be a lot of noise or was near the 25.0 814 \mss\ limit,
it was given a 0, possibly indicating that the image, in fact, is not
that of a real galaxy.  In the third category, location, the
galaxy's position in the mosaicked WFPC2 image was examined to determine the likelihood
that the `galaxy' was simply a region of higher than average surface
brightness within the  foreground LSB galaxy.
Again, a rating of 3 indicated the background galaxy was considerably removed from the
location of the foreground galaxy to make the possibility of it actually being a
part of the foreground galaxy extremely small.  A rating of 0 was given if, upon inspection
of the foreground galaxy it became clear the `background galaxy' was likely
a surface brightness enhancement within
the foreground galaxy (i.e it clearly lay within the foreground galaxy's nucleus or as a part
of a spiral arm).
Ratings of 1 and 2 were given to the galaxies which
lay in between these two extremes.
The three scores were then averaged and rounded to one
significant figure, and a final rating was given to the galaxies
(Column 5 in Table~\ref{tab:hstdat}).  Any galaxy which had an average score of 0
was dropped from the list, while a score of 1 indicates the identification of that
object as a background galaxy is questionable, a rating of 2 indicates that it
is likely the object is a background galaxy, and finally, a rating of 3
means the galaxy is clearly a background galaxy.

\section{The Data}

139 potential galaxies were identified in the four fields and surface photometry
was performed on each of these candidates.  All information
derived from the WFPC2 images on the galaxies is given in Table~\ref{tab:hstdat}
which is organized as follows:

\begin{list} {\setlength{\rightmargin 0.5in}{\leftmargin 0.5in}}
\item  {\bf Column 1:}  Galaxy names which correspond to our internal field sequencing
convention.  None of these galaxies have been previously identified.
\item  {\bf Column 2 \& 3:} RA and Dec of the galaxy, in J2000 coordinates, as found
using the STSDAS METRIC task.
\item  {\bf Column 4:} Galaxy type.
\item  {\bf Column 5:} Scorecard rating.
\item  {\bf Column 6:} The $\chi^2$ fit to the exponential profiles, assuming the
data has uniform error bars of \plm 0.05 \mss.  (As the fits were only carried
out from r = 0.2'' through r$_{25}$, this value is fairly accurate.)
\item  {\bf Column 7:} The $\chi^2$ fit to the r$^{1/4}$-type profiles, again
assuming the data has uniform error bars of \plm 0.05 \mss.
\item  {\bf Column 8:} The total integrated isophotal magnitude of the galaxy through the
814 filter.   Magnitudes are corrected for galactic extinction (treating
the 814 filter as a Johnson I band filter and the 300 filter as a Johnson
U band filter) but not for inclination or redshift (since that is
unknown).  Magnitudes are within 0.1 unless otherwise noted.
\item  {\bf Column 9:} Isophotal colors for the galaxies in 814 $-$ 300.  If
the galaxy couldn't be found in the 300 a minimum color is given, assuming the
galaxy would have been
detected if it had a minimum radius of 5 pixels and a brightness at least 2$\sigma$ above the
sky value.  Again, colors are  corrected for galactic extinction (treating
the 814 filter as a Johnson I band filter and the 300 filter as a Johnson
U band filter) but not for inclination of redshift.  Colors are within 0.2
magnitudes unless otherwise noted.
\item  {\bf Column 10:} The total integrated magnitude of the galaxy is calculated using
\begin{equation}\rm mag(\alpha )\:=\: \mu (0)\:-\: 2.5 log(2\pi \alpha^2)
\label{eq:malph}\end{equation}
where \alp\ is the exponential scale length in arcsec and \mo\ is the central
surface brightness in \mss.  If an exponential profile was not
fit to a particular galaxy's surface brightness profile this column is left blank.
\item  {\bf Column 11:} The central surface brightness of the galaxy in \mss,
when it was possible to obtain this from the profile fit.  Surface brightnesses are
within 0.1 \mss, unless otherwise noted.
In the case of r$^{1/4}$ profile galaxies, this is the effective surface brightness.
\item  {\bf Column 12:} The inclination corrected central surface brightness in \mss.
\begin{equation}\rm \mu_c (0)\:=\:\mu (0)\: -\: 2.5log(cos({\it i})), \end{equation}
where the inclination used is listed in column 11.
Note that this is a geometric path length correction which assumes no dust.
\item  {\bf Column 13:} The inclination angle (in degrees) as found by the
fitted ellipses
(equation~\ref{eq:inclin}).  The angle is accurate to \plm 5\degree.
\item  {\bf Column 14:} The exponential scale length in arcsec.  For the galaxies with
a r$^{1/4}$-type profile the value listed in this column is for R$_e$ (equation~\ref{eq:mur}).
\alp is not given for galaxies whose surface brightness profile is too irregular for a linear fit.
\alp (or r$^{1/4}$) are to within 0.1'' unless otherwise noted.
\item  {\bf Column 15:} The major axis radius in arcsec as measured at the \emu = 25.0 \mss\
isophote.  If the surface brightness profile errors exceeded 0.25 \mss\ before \emu
= 25.0 \mss, then the largest accurate radius is given.
r$_{25}$ is to within 0.1'' unless otherwise noted.
\end{list}
Comments about particularly interesting galaxies are contained in Appendix
A.

After identification of the background galaxies in the WFPC2 images, the Las
Campanas images were inspected to determine if any of these galaxies could
now be identified in the multi-filter ground based images in order to obtain
additional color information.
Twenty-seven
of the background galaxies were reliably identified in the ground based Las Campanas images, and
the colors of these galaxies through the B, V, and I filters were found.
The results are listed in Table~\ref{tab:gndback}.  Column 1 gives the
galaxy name, while column 2 gives the total integrated V magnitude for the galaxy.
Columns 3 and 4 give the B $-$ V and V $-$ I colors for the galaxies.  If a galaxy could not
be identified in one of the filters, a minimum color was found, assuming the
galaxy would have been identified were it at least 3 pixels in radius and had an
intensity 3$\sigma$ above the sky.  Finally, column 5 lists the radius
at which these colors were determined.  As with the WFPC2 images, the radii
chosen insured the errors were less than 0.25 mag.

\section{Surface Brightness Biases and Galaxy Types}

\subsection{The Early Type Galaxies}

Figures \ref{fig:hsta}, \ref{fig:hstb}, \ref{fig:hstc} are images
of selected galaxies in this survey that fall into one of our
structural categories.  Numerically, the majority of these
serendipitously discovered  galaxies (102 or 73\%) were fit with an
exponential profile, 27 of the galaxies (20\%) were fit by an
r$^{1/4}$-type profile, while the remaining 10 galaxies, or 7\%,
could not be fit by any profile, either because of their amorphous
appearance or  location in the frame.  Figure~\ref{fig:profs}
shows example surface brightness profiles for all categories. Our
overall result is thus fairly consistent with many previous
analyses of the Hubble Deep Field (i.e. Abraham, \etal 1994;
Abraham, \etal 1996; Driver, \etal 1995a, 1995b; Oemler, Dressler,
\& Butcher 1997).  This result will be discussed further in the
next section, and can be seen in Table~\ref{tab:surv}.

In Figure~\ref{fig:redshift} we plot effective radius (in arcsec) versus effective
surface brightness (in \mss) for our sample of  r$^{1/4}$  galaxies.  The
model running through the data shows the expected trend in these quantities
if we take a standard elliptical (R$_e$ = 5 kpc, M$_{B}$ = -21.0 (i.e. Kormendy 1977)
and redshift it,  it's angular size decreases according to
\begin{equation}\rm
\delta\:=\:{{R_e\:H_0\:q_0^2\:(1\:+\:z)^4}\over{zq_0\:+\:(q_0\:-\:1)\left({\sqrt{2q_0z\:+\:1}\:-\:1}\right)}}
\end{equation}
(Weinberg 1972).  Additionally, its surface brightness is dimmed
according to I$_e$(observed) = $\rm {{I_e}\over{(1+z)^4}}$, or equivalently
$\rm \mu_e(observed)\:=\:\mu_e\:+\:2.5log(1+z)^4$.
Figure~\ref{fig:redshift} shows the above plots for q$_0$ = 0.05, 0.55, \&
1.05.  Overlayed onto the plot are the points where the standard galaxy
would lie were it at a redshift of 0.5, 1.0, 1.5, 2.0, 2.5, \& 3.0 (left to right).
Although the data are limited, they do conform
well to this model indicating that r$^{1/4}$ galaxies do lie somewhere between
z = 0.5 and z=2.5, and begin to drop out of the
survey beyond a redshift of z$\sim$3.0 due primarily to the effects
of cosmological dimming.   While the variation around the model is
undoubtedly a result of the evolution of real stellar populations
(e.g. Schade, \etal 1999), the mean trend is consistent with the Tolman
test for the expansion of the Universe (see Wirth 1997;
Sandage \& Perelmutter 1990).
In this case, the large observed range in $\mu_e$ represents a large
range in redshift and not an intrinsic range in surface brightness perhaps
due to differences in star formation rates among these galaxies.
Note that there is one very deviant point in
this diagram which is both of large angular size and very low surface
brightness.   Spectroscopic follow-up of this galaxy
(m2-8) might be interesting.

\subsection{Recovering Freeman's Law}

The galaxies which fall into classification B (53 galaxies, or 38\%)
are those whose surface brightness profile is
{\it well fit}
by an exponential profile (equation~\ref{eq:mue}).  These galaxies have a median central
surface brightness of \lb \emo \gb = 22 \mss\ and a median scale
length of 0.7''.   Recall that the average surface brightness of the
background 814 sky was 23 \mss.  Hence, our median value of
\lb \emo \gb = 22 \mss\ is 1 \mss\ brighter than the sky background.
This is identical to selecting galaxies from the ground where the
Freeman value of 21.65 in the blue is approximately one magnitude
brighter than the blue sky background, as observed from the Earth.
This in fact, is the essence of the original argument of Disney (1976)
which was later quantified by Disney \& Phillips (1983) and
McGaugh \etal (1995).  The recovery of the 814 WFPC2 HST equivalent of
Freeman's law shows the remarkable uniformity of this selection effect
as applied to galaxy detection on any detector.  This, of course, means
that the serendipitous detection of true LSB galaxies with HST WFPC
will be as difficult as it has been from the ground (see Impey \&
Bothun 1997).

A few of these disk  galaxies show distinct
spiral structure, ranging from the sharp, well-defined spiral arms of 144-13,
through the very faint, yet still obvious arms of 284-2 and 731-1,
to the extremely clumpy, yet still visible arms of 734-13 and
283-2 (Figure~\ref{fig:hstb}).    Furthermore, about
one-third of these galaxies have significant luminosity excess in their
central regions and the rest are pure disk systems  or with
with central luminosity excesses too small to resolve.  
On the surface, this suggests that the
disk galaxies span a similar range in B/D ratio at z $\sim$ 1 as they
do at z=0, a point consistent with other studies (e.g. Wirth \etal
1994).  

\subsection {Bulge/Disk Deconvolution}

A feature of galaxy evolution, hitherto rather un-probed, is the
redshift evolution of the bulge-to-disk (B/D) ratio for disk galaxies.
This question is of key interest in understanding the origin of
S0 galaxies and whether or not they have always been present
or they are an evolutionary end product of normal astration processes
in disk galaxies (see Bothun \& Gregg 1990; Andreon 1998).   The identification
of relatively blue disk galaxies at moderate redshifts, which 
nonetheless have significant B/D, would indicate a population of disk galaxies
that likely do not have a long timescale for disk formation.  For
a reasonable star formation rate, the astration timescale in such
disks may well be half a Hubble time.  By z=0, such objects would
have the characteristics that define the SO class.   

Simard \etal\ (1999) have studied this issue by constructing a magnitude
versus size relation for a sample of distant, high surface brightness galaxies.
They find strong evidence that a wide range of bulge-to-total (B/T)
luminosities exist and that galaxies in their sample define regions
in the magnitude-size plane that are not occupied by local galaxies.
This suggests that significant evolution in B/T may well occur.  To
first order, evolution in B/T should also correlate with color evolution
but a large enough data set to look for this statistical signature
is not yet available.

In principal, our data is sufficiently deep that we can perform bulge/
disk deconvolutions on our surface brightness profiles.
Using the procedure of Schombert \& Bothun (1987) we have attempted to
derive B/D ratios for those exponential disks which show a clear
excess of light at small radii above the exponential.   In general,
this attempt was only made on galaxies where $r_{25}$ exceeded one arcsecond.
In many
of these cases, it was not possible to find an acceptable fit because
the excess light was not well described by an $r^{1/4}$ law.  This
is an important point.  Wirth \etal (1994) defined a light concentration
index as a means of quantitatively distinguishing early galaxies from
later galaxies.   However, the mapping of this light concentration index
onto a conventional bulge and disk structural scheme may be both complex
and ambiguous.  The excess light, or the light which causes a high value
for the light concentration index could well be a bulge, or an extended
nuclear starburst, or another exponential disk with a short scale length.
Indeed, in our sample we have found examples of these so called
``exponential bulges'' (e.g. Carollo 1999; Seigar \& James 1998;
Moriondo \etal\ 1998) in which the composite profile is best fit as the
sum of two exponential disks.  

Approximately 25\% of our sample exhibited a surface brightness profile
morphology that made them candidates for B/D deconvolution.  In most
cases, the deconvolution failed to converge, again largely because
the excess light about the exponential disk could not be well fit by
an $r^{1/4}$ law.  In some cases, the deconvolution failed because the
disk fit was too noisy.  However, we were able to adequately fit 7 profiles
with a standard bulge+disk.  These profile fits are shown in Figure~\ref{fig:bdfit}
and the fitting parameters are shown in Table~\ref{tab:bdfit}.  It is clear from
inspection of the figures that the bulge+disk fits are not particularly
good (in the strict $chi^2$ sense) but these are the only 7 objects
for which such a fit was even approximately decent.  While it would be
silly to generalize on the basis of only 7 galaxies, we do note that
these B/D deconvolutions define a fairly large range in B/D ratio and
include several objects which do have large B/D.   We estimate the
uncertainties on our formal values of B/D at $\pm$ 30\%.

This large range in B/D
is consistent with the large range in B/T seen in the Simard \etal
(1999) sample.  The overall low frequency, however, at which a bulge+disk
fit could be found, indicates that, relative to nearby galaxies, these
distant galaxies are structurally noisy.  That in itself
may be an important conclusion from this study.  That is, the vast
majority of the detected, presumably distant, galaxies have a luminosity
distribution that can not be deconvolved into simple bulge + disk
components.  

Finally, we call attention to the case of 283-10 as an example of
a structurally ambiguous object (see also Andreon 1998).   In the
top panel of Figure~\ref{fig:283-10} we show the surface brightness profile with
an exponential disk fitted to the outer regions.  The exponential
fit is reasonably good and the clear excess luminosity at r $<$ 1.0''
indicates a bulge component.   The middle panel shows the
best fitting bulge+disk model.  The overall fit is not very good and
the resulting B/D ratio is $\sim$ 2.0 $\pm$ 0.2.  This is a high
value of any disk system and, if real, would be a good example of
an S0 galaxy, similar to say NGC 7814 (see Bothun \etal 1992) or
NGC 5866 both of which have approximately a 25\% contribution to the
total V-band luminosity from the disk component.  But, should we
believe this deconvolution?  In the bottom panel we show a pure
$r^{1/4}$ fit on the data.  The fit is not great but most of the
deviation is occurring at radii larger than 2.0 arcseconds.  So
what is the nature of 283-10?  Is it a structurally noisy elliptical
galaxy?  Does it have a small disk and is therefore an S0?  Or is it
a disk dominated system with a strong nuclear excess of light that
is not related to a bulge?  Clearly, the answer is ambiguous and this
one object is a strong testimony to the difficulty of performing
accurate structural analysis of distant galaxies.

\subsection {Structurally Noisy Galaxies}

As is well known (i.e. Abraham, \etal 1994; Abraham, \etal 1996; Driver, \etal 1995a, 1995b;
Oemler, Dressler, \& Butcher 1997, Smail \etal 1999) the frequency of galaxies with ``irregular'',
``amorphous'', or ``peculiar'' appearance seems to rapidly
increase with redshift.  Similar results are seen here to the extent
that we can equate the 49 galaxies which have surface brightness profiles too
irregular to classify  (e.g. Category C) with morphologically peculiar
objects.  These galaxies typically exhibit a fairly clumpy appearance,
perhaps indicative of some kind of asymmetric star formation activity/
dust distribution (e.g. Smail \etal 1999) or tidal
encounter with another galaxy (see Patton \etal 1997).
Indeed most deep galaxy surveys reveal an unusually high
number of galaxies at z \gta 0.5 to be interacting (i.e. Oemler,
Dressler, \& Butcher 1997; Pascarelle, \etal 1996; Abraham, \etal 1996; Driver,
\etal 1995a).  The probability that at least a few of the
galaxies in this group have experienced a merger is fairly high and
some examples of this appear as overlayed galaxies in our images.    For
instance, of the galaxies in category C that are large enough to
produce reasonable surface photometry, most exhibit exponential surface
brightness profiles with small overlayed `bumps' which could be individual
regions of star formation, perhaps triggered by a merger or strong
interaction.     Interestingly, the median central surface brightness
for category C galaxies (\lb \emo \gb = 22.7 \mss) is lower than category
B galaxies and hence they should be selected against.  However,
their lumpy appearance (e.g. surface brightness
enhancements) greatly aids in their visual detection.  This suggests that part
of the reason for the apparent increase in galaxies of ``peculiar''
morphology is simply they are easier to recognize against the background
sky noise than galaxies with more smooth appearance but lower than
average surface brightness.

\subsection{Inclination, Size, Colors and Redshift}

The median inclination of the galaxies in this survey is 50\degree, close to
the expected value of 60\degree in random phase space.  The distribution
of inclination is shown in (Figure~\ref{fig:hsti}).
A significant fraction of the galaxies (39, or 28\%) have {\it i} \gta 65\degree.
These galaxies would fall into Cowie, \etal's (1995) classification of `chain galaxies,'
since many also show the bright knots (of star formation?)
 inherent in the chain galaxy
classification.  Two good examples of this phenomenon are 734-9
and 284-15.  734-9 is an edge-on galaxy, with {\it i} = 75\degree.
In addition to its well defined central bulge, 734-9 clearly shows
a number of the knots of star formation discussed in Cowie, \etal.
Due to their placement and the clear central bulge of the galaxy,
these `knots' appear to be relatively transparent, edge-on spiral
arms.  Each galaxy in the apparent galaxy group that creates
284-15 is also fairly edge-on, ranging from 58\degree \lta\ {\it i}
\lta\ 77\degree.  The close proximity and edge-on nature of these
three galaxies indicates they are interacting, and are possibly
simply the bright regions of the same galaxy. Because the same
localized spots occur in the galaxies with {\it i} \lta\ 65\degree,
and because many local galaxies have been found with inclinations
equally as high (e.g. Dalcanton \& Schectman 1996; O'Neil, Bothun,
\& Cornell 1997a), it is likely these galaxies do not belong to a
new galaxy classification but instead are the same (albeit more
inclined) as the other galaxies in this survey.

Lacking the detailed color information to properly infer photometric
redshifts (e.g. Connolly, Szalay, \& Brunner 1998,),  we can only
estimate redshifts by using the measured scale lengths.
The average scale length is \lb \alp \gb = 0.7'', with a range from
0.1'' \lta\ \alp \lta\ 3.1''.  Assuming these are typical disk galaxies, with
1 kpc \lta\ \alp \lta\ 5 kpc
(H$_0$ = 100 km/s/Mpc) (van der Kruit 1987; Gr\o sbol 1985), gives a  probable
redshift range of 0.2 -- 1.0 for q$_0$ = 0.05 -- 1.0.

Figures~\ref{fig:bvvi} and \ref{fig:alphavi} show the two color
diagram for the background galaxies that could also be identified
in the ground-based images and the relation between measured scale
length and V$-$I color. Comparing these colors with those from
galaxies with known redshifts in the Hubble Deep Field indicates
the background galaxies have redshifts lying between 0.5 \lta\ z
\lta\ 1.5, though the possibility exists, from this comparison,
that the galaxies lie considerably farther away (i.e. Phillips
\etal 1997; Lowenthal \etal 1997; Madau \etal 1996).  However, in
this case their disk scale lengths would exceed 5 kpc. The three
galaxies with B$-$V \lta\ 0.55 and V$-$I \lta\ 1.0  all have
r$_{25}$ values of 1 -- 1.5'' and are most probably low luminosity
irregular galaxies at relatively low redshift.  The majority of
points, however, are of small angular size and red color (with one
prominent exception), suggesting a probable redshift of 0.5 \lta\ z
\lta\ 1.5. To first order, the relatively noisy correlation between
observed angular scale length and color does suggest that the
redder objects are simply farther away. This would place them at
the distance of many of the galaxies identified in the medium-deep
HST WFPC2 surveys and, as will be discussed in the next section,
the morphology of these galaxies is similar to that of the
WFPC/WFPC2 deep and medium-deep surveys.

\section{Comparison with Other WFPC2 Surveys}

Extensive work has been done to morphologically classify
the galaxies found in both the deep (HDF) and medium-deep (MDS) WFPC2 surveys
(i.e. Marleau \& Simard 1998; Oemler, Dressler, \& Butcher 1997;
Driver, \etal 1995a; Driver, \etal 1995b; Griffiths, \etal 1994).
On average, the morphological mix of galaxies for both the HDF and MDS is
given as: A type galaxies (E/S0) 16\% - 41\%; B type galaxies (Sabc) 31\% - 53\%;
C type galaxies (Sd/Irr/Pec) 15\% - 47\% (Table~\ref{tab:surv}).
Considering only those galaxies for which
a surface brightness profile could be found, the galaxies in this survey
are distributed as 21\% A type, 41\% B type and 38\% C type, a distribution
similar to the majority of HDF and MDS classification schemes.
Again, we emphasize that our classification scheme is based strictly on
profile fit to surface photometry.    The issue now becomes whether
this distribution of A, B and C types is representative or whether the
results are driven by volume selection effects.   If Figures~\ref{fig:redshift} and \ref{fig:alphavi}
 are indirect
volume sampling indicators then it seems clear that unless the disk
galaxies selected here have intrinsically large scale length, the volume
involved selecting type A galaxies is considerably larger than selecting
type B galaxies.  This strongly suggest that the absolute volume density
of type A galaxies, as observed outside of clusters, is quite low in
comparison to that of type B galaxies.  If this is correct, then this
situation seems to be unchanged with respect to what we observe at z=0.

Support for this view comes from the results of other surveys which show that
the morphological mix deduced in a survey does seem to be dependent
on the magnitude limit of that survey.  This, of course, is the selection
function and its currently unclear if going to fainter limits means
more full sampling of the galaxy luminosity function (and its evolution
with time) or going to a large volume (hence longer look back times).
Spectroscopic follow-up surveys (e.g. Morris \etal 1998) tend to
be more consistent with the volume effect although this may be a direct
reflection of surface brightness bias in the spectroscopy.
The morphological
mix listed above is from surveys with limits of m$_I\:\sim$ 21.75 or brighter, while
the galaxies in our survey have a limit of $\sim$ 24, with a significant portion
lying between 22 \lta\ m$_I$ \lta\ 24.  Driver, \etal (1995b) analyzed the morphological
mix of galaxies in the HDF with 23.0 \lta\ m$_I$ \lta\ 24.5, and show that the
percentage of E/S0 galaxies in this sample is considerably lower (16\%) than
in the brighter sample, with most of the difference lying in the Sd/Irr category (47\%).
Thus it's likely that any difference between our sample and previous
WFPC2 galaxy surveys lies in the fact that our sample is considerably fainter than
previous studies.  As disk galaxies are more prevalent at lower absolute
magnitudes, this would be consistent with our technique selecting objects
further down the luminosity function.

In contrast,
using an automated, 2-dimensional photometric decomposition algorithm, Marleau \& Simard
(1998, MS) analyzed galaxies in HDF down to m$_{814}$ = 26.0 (m$_I\:\sim$ 24.5)
and found only 8\% of the galaxies to be bulge dominated (A type).  The first
reason given by MS to explain the discrepancy between the morphological
classifications in their survey and that of previous HDF morphological classifications
is that the subjective nature of the visual classification used by most groups
analyzing the HDF data biases towards finding more early-type galaxies
(this, of course, is a surface brightness selection bias).
This bias within the visual classification system effects primarily small (r \lta\ 0.31'')
round galaxies which are typically labeled as E/S0 galaxies.  As the algorithm
run by MS does not have such biases it found a considerably lower percentage
of A type galaxies.  On the other hand, our classification system is 
structurally based and hence similar to that used by MS and we do not
find such a low percentage of early type galaxies.
The second argument put forth by MS is that
many of the galaxies in their survey lay at z \gta 1, considerably farther than
the majority of the galaxies examined by other groups.   However, it is
unclear if MS are viewing morphological evolution in galaxies as being
responsible for their low percentage of early types.  If, instead, the faintest
galaxies observed by MS, are in fact not at z \gta 1 but are simply
low luminosity galaxies, then the 8\% number they find is quite similar
to what is found in the nearby Universe, if you sample far enough down
the galaxy luminosity function.

\section{Conclusions}

While using the HST WFPC2 to image four \lsbgs\ in the direction of the Virgo
cluster we discovered 139 potential background galaxies shining through the LSB galaxies.
We performed surface photometry on each of these images
and classified them into various structural types depending on the form
of the surface brightness profile.  Our overall results are the following:

1.  The combination of angular sizes and limited color information is
consistent with these galaxies occupying the redshift range
0.5 \lta\ z \lta\ 2.5.  This places the galaxies at the same distance
as many of the galaxies discovered in the Hubble Deep Field and the Medium Deep Survey.

2.  The value of \Imo found for the ``disk'' galaxies is approximately
one magnitude brighter than the F814W sky background.  Although the detector is
much different, this is a manifestation of the kind of surface brightness
selection bias that leads to Freeman's Law - i.e. values of \Bmo that
are approximately one mag brighter than the terrestrial blue sky background.

3.  The percentage of $r^{1/4}$ galaxies found in this survey is similar
to that found by others (e.g. $\sim$ 28\%).  There are, however, indirect
suggestions in the data that the volume over which $r^{1/4}$ galaxies are
selected is significantly larger than the volume over which disk galaxies
are selected.   This suggests that the space density of $r^{1/4}$
galaxies in the general field from z =0 to z $\sim$ 2.5 is low;
that is, this population may be largely confined to galaxy clusters.

4.  Rather few of the galaxies can be reliably deconvolved into a bulge
+ disk.  Even those that could did not have particularly good fits.
To first order, this indicates that the structural components
of disk galaxies have not fully formed at these redshifts.  Interestingly,
the small sample of bulge+disk fits that we did obtain show a large
variation in the bulge-to-total luminosity ratio and we did detect some
disk galaxies with significant B/D.  Such objects could be visually
classified as ellipticals.  We have also highlighted one object, 283-10,
as an excellent example of how difficult the classification exercise
can be.

5.  It seems clear that surface brightness information, when coupled
with broad-band colors, can help to better quantify the rate of
``morphological'' evolution of galaxies.  However, the recovery of
Freeman's Law from this data, together with the known cosmological
dimming effect, (1+z)$^4$, means that the biases against selecting
intrinsically LSB objects at high redshift are severe.   There may
well be large numbers of such objects, as is the case at z=0 (e.g.
O'Neil \& Bothun 2000), that simply can easily escape detection.
This demands that caution must be taken when using the number  density
of galaxies, as a function of redshift, as a cosmological probe.

Support for this work was provided by NASA through grant number GO-05496
from the Space Telescope Science Institute, which is operated by the
Association of Universities for Research in Astronomy, Inc, under NASA
contract NAS5-26555.  We further acknowledge support from the NSF

\section{Appendix A: Few Good Galaxies}

Having found 139 previously uncatalogued galaxies, we cannot go into
detail about each one.  A few of the galaxies, though, are worthy of mention,
due to their unusual shape or surface brightness
profile. Images of these, and all the galaxies in this survey, can be found at
http://guernsey.uoregon.edu/$\sim$karen.  These unusual galaxies are listed below:
\begin{list} {\setlength{\rightmargin 0.5in}{\leftmargin 0.5in}}
\item  {\bf 731-5:} This appears to be two galaxies nearing the end of their merger, with
one galaxy (the chosen core in the surface brightness profile) considerably brighter than the other.
Both galaxies likely had well-formed cores before the merger
began.
\item  {\bf 731-6:} This galaxies is fairly small and diffuse, and
was only found due to its bright central core (\emo $\sim$ 21.5 \mss),
The galaxy lies in a particularly noisy region of the image and therefore is difficult to
classify.
\item {\bf 732-3:}  This is a fascinating galaxy, having a clear spiral core surrounded
by a diffuse halo which has its own spiral arm (Figure~\ref{fig:hstc}a).
\item  {\bf 733-10:} The core (peak intensity) of 733-10 lies far from the galaxy's physical
center.  Although it is, of course, impossible to know the cause of this, the
fact the core is highly non-spherical lends credence to the conclusion that this
galaxy has recently experience a strong tidal influence, either by a passing galaxy
or by a galaxy which has recently merged with 733-10.
\item  {\bf 734-10:}  These highly elliptical ({\it i} = 80\degree) galaxies may be
interacting, though their ellipticity makes that determination difficult.
\item  {\bf 283-3:} This object
 has a star overlaying the galaxy between r = 0.7"  ad
1.3".  The star was masked in the surface brightness profile, leaving  an
artificial depression.
\item  {\bf 283-9, 283-10:} This pair of galaxies consists of a large, bright, and presumably
elliptical galaxy (283-10) and a much small galaxy.  The excess of gas
between the galaxies makes it appear as if they're interacting tidally.  Due to the large
discrepancy in size between the two galaxies, it is likely 283-9 will be, or is being, ripped apart by, or pulled
into 283-10.
\item  {\bf 284-7:} 284-7 has an usual, V-shaped morphology, with a bright central core.  This
galaxy may actually be two small overlapping galaxies, or we could be seeing star forming
knots embedded in a larger, but much fainter, galaxy.
\item  {\bf 284-15:}  Based on their close proximity to each other,
this system appears to be three separate, interacting
galaxies.  All three
galaxies are highly inclined ({\it i} = 58\degree, 77\degree, 69\degree, respectively), and all
have remarkably similar position angles.
\item  {\bf 142-8:} This appears to be two (or three) galaxies interacting galaxies.  142-8
contains two bright cores and has a highly elliptical appearance ({\it i} = 73.3\degree).  It may be
the result of a recent merger, or the non-centralized knot of star formation
may be due to the close proximity of its companion.
\item  {\bf m2-21:}  At r$\rm _{major}$ = 0.85", m2-16 has two bright circular regions.  These
may be foreground stars, or even a foreground galaxy, or they may be star forming regions in
the galaxies.  If the spots are regions of heightened star formation,
then, because they are far from the galaxy's peak intensity,
they were most likely externally triggered.
\item  {\bf m3-9:} This galaxy has previously been given a stellar classification due to its
round shape.  The high resolution of the WFPC2, however, enabled us to view the galaxy's
faint disk and thereby determine its true nature as a galaxy.
\end{list}

The four images which constitute the full field of the WFPC2 are strongly vignetted in the lower
columns and rows due to the aberration of the primary beam dividing the light from sources near
these edges.  The increased noise affects the faint end of the surface brightness profiles of the
following galaxies:
284-8, 284-24, 733-1, 733-14, 734-1, 734-2, 734-8, 141-5, 143-15 and 143-16,
m2-30, m4-1 and m4-2.

\clearpage \centerline{\bf References}

Abraham, R..G. \etal 1996, ApJS, 107, 1\\

Abraham, R..G. Valdes, F., Yee, H., van den Bergh, A.  1994, ApJ, 432, 75\\

Andreon, S. 1998 ApJ 501,533\\

Barger, A.J. \etal 1998 AJ, in press\\

Balogh \etal 1999, preprint\\

Bothun, G., \etal 1992 PASP 104, 1220\\

Bothun, G., \& Gregg, M. 1990 ApJ 350, 83\\

Carollo, C. M. 2999 ApJ 523, 566\\

Cohen, J, \etal 1996 ApJ 471, 5\\

Connally, A.J., Szalay, A.S., \& Brunner, R.J. 1998 ApJ 499, 125\\

Cowie, Lennox L., Songaila, Antoinette, Hu, Ester M., \& Cohen, J. G. 1996 AJ 112, 839\\

Cowie, L., Hu, E., \& Songalia, A.  1995, AJ, 110, 1576\\

Dalcanton, J. \& Schectman, S. 1996, ApJ, 465, 9\\

Disney, M., \& Phillipps, S.  1983 MNRAS 205, 1253\\

Disney, M. 1976 Nature 263, 573\\

Driver, S., \etal 1995a, ApJ, 453, 48\\

Driver, S., \etal 1995b, ApJL, 449, 23\\

Ellis, R., \etal 1996 MNRAS, 280, 235\\

Griffiths, R.E., \etal 1994 ApJ 437, 67\\

Gr\o sbol, P. 1985 A\&AS 60, 261\\

Hammer, F, \etal 1997 ApJ 481, 49\\

Hogg, D., \etal 1999, preprint\\

Impey, C., \& Bothun, G. 1997 ARA\&A 35, 267\\

Impey, C., Bothun, G., Malin, D.  1988, ApJ, 330, 634 (IBM)\\

Kormendy, John 1977 ApJ 218, 333\\

Krist, J. 1996, {\it Tiny Tim Manual} http://scivax.stsci.edu/krist/tinytim.html\\

Lilly, S. J., Le Fevre, O., Crampton, David, Hammer, F., \& Tresse, L. 1995 ApJ 455, 50\\

Lowenthal, J. 1997 ApJ 481, 673\\

Madau, Piero, 1997 IAUS 186, 188\\

Madau, Piero, \etal 1996 MNRAS 283, 1388\\

Marleau, F. R. \& Simard, L. 1998 AAS 192, 2503 (MS)\\

McGaugh, S., Bothun, G., \& Schombert, J. 1995 AJ 109, 2019\\

Meurer, G., Heckman, T., Lehnert, M., Leitherer, C., \& Lowenthal, J. 1997 AJ 114, 54\\

Moriondo, G. \etal 1998 A\&AS 130, 8\\

Morris, S. \etal 1998 ApJ, in press\\

Oemler, A., Dressler, A., \& Butcher, H  1997, ApJ, 474, 561\\

O'Neil, K. Bothun, \& Impey 1999, ApJ, preprint\\

O'Neil, Bothun, Impey, \& McGaugh 1998, AJ, 116, 657\\

O'Neil, K, Bothun, G. \& Cornell, M. 1997a, AJ, 114, 2448\\

O'Neil, K, Bothun, G. Schombert, J., Cornell, M., Impey, C. 1997b, AJ, 113, 1212\\

O'Neil, K.  1997, Ph.d. dissertation, University of Oregon, Eugene\\

Pascarelle, S. M., Windhorst, R. A., Driver, S. P., Ostrander, E. J., \& Keel, W. C.
1996 ApJ 456, 21\\

Patton, D. \etal 1997 ApJ 475, 29\\

Phillips, A., \etal 1997 ApJ 489, 543\\

Ratnatunga, \etal 1999 AJ 118, 86\\

Remy, M. \etal 1997 {\it 1997 HST Calibration Workshop} S. Casertano, \etal eds.\\

Sandage, A., \& Perelmuter, J-M. 1990, ApJ, 361, 1\\

Schade, \etal  1999, preprint\\

Schombert, J. \& Bothun, G. 1987 AJ 93, 60\\

Seigar, M. \& James, P. 1998 MNRAS 299 672\\

Silva, D. \& Bothun, G.  1998 AJ 116, 85\\

Simard, L., \etal 1999 ApJ 519, 563\\

Smail, \etal 1999, preprint\\

Steidel, C., \etal 1999, preprint\\

Steidel, C., Giavalisco, M., Pettini, M., Dickinson, M., Adelberger, K, 1996 ApJ 462, 17\\

Thomsen, B., \& Frendsen, S. 1983, AJ, 88, 789\\

Van Der Kruit, P. C. 1987 A\&A 173, 59\\

von Dokkum, P., \etal 1998 ApJ in press\\

Weinberg, S. 1972 {\it Gravitation and Cosmology: The Principles and Applications of the
General Theory of Relativity} John Wiley \& Sons: New York\\

Wirth, G., Koo, D., \& Kron, D. 1994 ApJ 432, 464\\

Wirth, G., \etal 1997 PASP 109, 344\\
\clearpage
\centerline{\bf FIGURES}

\figcaption[obi_backg.fig1.ps]{HST WFPC2 mosaicked images of V1L4 (a), V2L8 (b), V7L3 (c) and Malin 1 (d)
taken through the 814 (I band) filter.  The images are 2.6 arcminutes across.
\label{fig:wfpc}}

\figcaption[obi_backg.fig2.ps]{Representative sample of surface brightness profiles for the three
categories.  Figure~\ref{fig:profs}{\it a} are all category A,  Figure~\ref{fig:profs}{\it b}
are category B, and  Figure~\ref{fig:profs}{\it c} are category C.
\label{fig:profs}}

\figcaption[obi_backg.fig3.ps]{Examples of some of the more
interesting (morphologically) background galaxies discovered
during our study.  The galaxies shown are 732-3, 733-10, 282-3,
and 142-31 (a - d, respectively). \label{fig:hsta}}

\figcaption[obi_backg.fig4.ps]{Examples of spiral background
galaxies discovered during our study. Images a - e are 731-1,
284-2, 144-13,  283-2, and 734-13, respectively. \label{fig:hstb}}

\figcaption[obi_backg.fig5.ps]{Examples of merging background
galaxies discovered during our study. Figure~\ref{fig:hstc}a shows
731-5, while Figure~\ref{fig:hstc}b shows 734-10.  In
Figure~\ref{fig:hstc}c is 283-9 at the top, and 283-10 at the
bottom. Figure~\ref{fig:hstc}d shows the three galaxies which
comprise 284-15. \label{fig:hstc}}

\figcaption[obi_backg.fig6.ps]{A plot of the effective radius (in
'') versus $\mu_{e_{814}}$ (in \mss) for the sample of  r$^{1/4}$
galaxies.  The lines are for q$_0$ =0.05 (solid line), 0.55
(dashed line), and 1.05 (dash-dotted line), the circles are the
actual data, and the crosses are for the standard galaxy
(R$_e$=5.0 kpc, $\mu_{e_{814}}$ 21.0 \mss) with z = 0.5, 1.0, 1.5,
2.0, 2.5, \& 3.0 (left to right). (See Sandage \& Perelmuter 1990,
Thomsen \& Frandsen 1983 for more information on this procedure.)
\label{fig:redshift}}

\figcaption[obi_backg.fig7.ps]{Surface brightness profiles
for the 7 galaxies which were successfully fit with a standard
bulge + disk profile.\label{fig:bdfit}}

\figcaption[obi_backg.fig8.ps]{Surface brightness profile of
283-10 with a pure exponential fit ({\it a}), a bulge + disk fit
({\it b}), and a pure r$^{1/4}$ fit ({\it c}).\label{fig:283-10}}

\figcaption[obi_backg.fig9.ps]{Histogram showing the inclination (in degrees) for all the galaxies in
this survey.
\label{fig:hsti}}

\figcaption[obi_backg.fig10.ps]{Two color diagram for the background galaxies that could also be identified
in the ground-based images.
\label{fig:bvvi}}

\figcaption[obi_backg.fig11.ps]{Scale length (in '') vs. V$-$I color for the background galaxies that
could also be identified in the ground-based images.
\label{fig:alphavi}}

\clearpage
\centerline{\bf TABLES}

Table~\ref{tab:hstf}. J2000 Coordinates for the Surveyed Fields.

Table~\ref{tab:apmag}. Aperture Magnitudes.

Table~\ref{tab:hstdat}.  Photometry for All Galaxies in this Study.

Table~\ref{tab:gndback}.  Colors of the 29 Galaxies Detected on the Ground.

Table~\ref{tab:surv}.  Comparison of the Galaxy Types for Different Surveys.

Table~\ref{tab:bdfit}.  Bulge + Disk Fitting Parameters.

\clearpage

\begin{table}[p]
\caption{\label{tab:hstf}}
\vskip 0.2in
\begin {center}
\begin{tabular}{lccc}
\hline\hline
\multicolumn{4}{c}{}\\
{\bf Field}& {\bf RA}&{\bf Dec}&{\bf Date}\\
\multicolumn{4}{c}{}\\
\hline
\multicolumn{4}{c}{}\\
V2L8& 12:31:27.1& 14:37:14.1& 01/05/96\\
V7L3& 12:29:06.2& 12:53:34.7& 03/08/97\\
V1L4& 12:34:52.3& 14:12:21.3& 03/10/96\\
Malin 1& 12:37:08.8& 14:18:45.9& 03/10/96\\
\multicolumn{4}{c}{}\\
\hline \hline
\end{tabular}
\end{center}\end{table}

\begin{table}[ht]
\dummytable{\label{tab:apmag}}
\end{table}

\begin{table}[ht]
\dummytable{\label{tab:hstdat}}
\end{table}

\begin{table}[ht]
\dummytable{\label{tab:gndback}}
\end{table}

\begin{table}[ht]
\dummytable{\label{tab:surv}}
\end{table}

\begin{table}[ht]
\dummytable{\label{tab:bdfit}}
\end{table}

\end{document}